\pdfoutput=1

\documentclass[twocolumn,final]{svjour3}

\smartqed

\usepackage{graphicx,amssymb}
\usepackage{epsf,psfig}

\journalname{Nonlinear Dynamics}

\begin{document}

\title{Application of new dynamical spectra of orbits in Hamiltonian systems}

\author{Euaggelos E. Zotos}

\institute{Euaggelos E. Zotos:
\at Department of Physics, \\
Section of Astrophysics, Astronomy and Mechanics, \\
Aristotle University of Thessaloniki \\
GR-541 24, Thessaloniki, Greece \\
email:{evzotos@astro.auth.gr}
}

\date{Received: 1 March 2012 / Accepted: 21 March 2012 / Published online: 19 April 2012}

\titlerunning{Application of new dynamical spectra of orbits in Hamiltonian systems}

\authorrunning{Euaggelos E. Zotos}

\maketitle

\begin{abstract}

In the present article, we investigate the properties of motion in Hamiltonian systems of two and three degrees of freedom, using the distribution of the values of two new dynamical parameters. The distribution functions of the new parameters, define the $S(g)$ and the $S(w)$ dynamical spectra. The first spectrum definition, that is the $S(g)$ spectrum, will be applied in a Hamiltonian system of two degrees of freedom (2D), while the $S(w)$ dynamical spectrum will be deployed in a Hamiltonian system of three degrees of freedom (3D). Both Hamiltonian systems, describe a very interesting dynamical system which displays a large variety of resonant orbits, different chaotic components and also several sticky regions. We test and prove the efficiency and the reliability of these new dynamical spectra, in detecting tiny ordered domains embedded in the chaotic sea, corresponding to complicated resonant orbits of higher multiplicity. The results of our extensive numerical calculations, suggest that both dynamical spectra are fast and reliable discriminants between different types of orbits in Hamiltonian systems, while requiring very short computation time in order to provide solid and conclusive evidence regarding the nature of an orbit. Furthermore, we establish numerical criteria in order to quantify the results obtained from our new dynamical spectra. A comparison to other previously used dynamical indicators, reveals the leading role of the new spectra.


\end{abstract}

\section{Introduction}

Over the last half century, a large number of studies have been dedicated to efforts on finding new ways to characterize the nature of orbits in Hamiltonian systems [1,2,7,18,22,26,30]. Among them the dynamical spectra (i.e. distribution of the values of a given parameter along the orbit) have been proved a very reliable tool for the exploration of the properties of motion in dynamical systems. In previous works [37] and [11-12], presented a new method that allows the distinction between domains of ordered and chaotic motion in Hamiltonian systems of two and three degrees of freedom, or in 2D and 4D symplectic maps. This method is based on the calculation of stretching numbers, helicity angles and twist angles (see the above referenced papers for theoretical details) and their respective dynamical spectra.

Consider the Poincar\'{e} section map of a Hamiltonian dynamical system. Let $\xi_{i+1}$ be the next image of an infinitesimal deviation $\xi_i$ between two nearby orbits. The ``stretching number" $\alpha_i$ [37], or the ``local Lyapunov indicator" [16] is defined as
\begin{equation}
\alpha_i = \ln \left|\frac{\xi_{i+1}}{\xi_i}\right|,
\end{equation}
i.e. it is a 1-step Lyapunov Characteristic Number (LCN). The spectrum of stretching numbers is their distribution function, namely
\begin{equation}
S(\alpha) = \frac{\Delta N(\alpha)}{N \Delta \alpha},
\end{equation}
where $\Delta N(\alpha)$ is the number of stretching numbers in the interval $\left(\alpha, \alpha + \Delta \alpha \right)$, after $N$ iterations. Other types of dynamical spectra have also been used in previous studies [3,5,6,9,10,19,20,28].

In the present paper we will demonstrate and prove the high efficiency of the new dynamical spectra in separating ordered from chaotic domains in a realistic galactic model. The separation can be obtained even with an integration time for individual orbits, as short as 5000 time units. Thus, our new dynamical indicators are faster than other spectral methods, e.g. the frequency analysis method [21], and much more faster than the calculation of the Lyapunov Characteristic Exponent (LCE) [23]. In particular, our new dynamical spectra are very fast and practical tools for locating multiple islands of invariant curves embedded in the chaotic sea, which correspond to resonant orbits of higher multiplicity. In addition, they can be used to provide a quick estimation of the fraction of the phase space volume occupied by ordered or chaotic orbits, at many different energy levels. This information is needed, when one wishes to construct self-consistent models, in which many different orbits must be combined.

We shall test our new dynamical spectra in the potential
\begin{equation}
V_{tot}(x,y,z) = V_{har}(x,y,z) + V_{nuc}(x,y,z),
\end{equation}
where
\begin{equation}
V_{har}(x,y,z) = \frac{\omega ^2}{2}\left(x^2 + a y^2 + b z^2 \right)
\end{equation}
and
\begin{equation}
V_{nuc}(x,y,z) = - \frac{M_n}{\sqrt{x^2 + y^2 + z^2 + c_n^2}}.
\end{equation}
Here $a$ and $b$ are the flattening parameters, $M_n$ and $c_n$ is the mass and the scale length of the nucleus respectively, while the parameter $\omega$ is used for the consistency of the galactic units. We use a system of galactic units, where the unit of mass is $2.325 \times 10^7$ M$_\odot$, the unit of length is 1 kpc and the unit of time is $0.997748 \times 10^8$ yr. The velocity unit is 10 km/s, while $G$ is equal to unity. In the above units we use the values: $\omega$ = 10 km s$^{-1}$kpc$^{-1}$, $a=4, b=1.25, M_n=50$, while $c_n=0.25$. The 3D gravitational potential $V_{tot}$, describes satisfactorily the central parts of an elliptical galaxy and consists of two parts: the first part, that is $V_{har}$ is the potential of a 3D anisotropic harmonic oscillator, derived from Taylor expansion of the logarithmic potential, while the second part, $V_{nuc}$, is a Plummer potential of a spherical nucleus. This particular dynamical system has been studied in detail in [38] (hereafter Paper I). The main reason which justifies the choice of the potential (3) is that it displays a large variety of resonant orbits, different chaotic components and also several sticky regions. Therefore, it provide us an excellent opportunity in order to test and also reveal the great efficiency of our new dynamical parameters.

The Hamiltonian corresponds to the potential (3) is
\begin{equation}
H = \frac{1}{2} \left(p_x^2 + p_y^2 + p_z^2 \right) + V_{tot}(x,y,z) = h,
\end{equation}
where $p_x, p_y$ and $p_z$ are the momenta per unit mass conjugate to $x, y$ and $z$ respectively, while $h$ is the numerical value of the Hamiltonian. All the outcomes of the present research, are based on the numerical integration of the equations of motion
\begin{eqnarray}
\ddot{x} &=& -\frac{\partial \ V_{tot}}{\partial x} = - \left[\omega^2 + \frac{M_n}{\left(x^2 + y^2 + z^2 + c_n^2\right)^{3/2}}\right] x, \nonumber\\
\ddot{y} &=& -\frac{\partial \ V_{tot}}{\partial y} = - \left[a \omega^2 + \frac{M_n}{\left(x^2 + y^2 + z^2 + c_n^2\right)^{3/2}}\right] y, \nonumber\\
\ddot{z} &=& -\frac{\partial \ V_{tot}}{\partial z} = - \left[b \omega^2 + \frac{M_n}{\left(x^2 + y^2 + z^2 + c_n^2\right)^{3/2}}\right] z,
\end{eqnarray}
where the dot indicates derivative with respect to the time. We integrated the equations of motion with a double precision Runge-Kutta algorithm of 7 - 8$^{th}$ order (RK78), in Fortran 77 [31]. The accuracy of the outcomes, was checked by constancy of the energy integral (6), which was conserved up to the fifteenth significant figure.

Our aim is to test and explore the properties and also the advantages of our new dynamical spectra. The $S(g)$ spectrum will be applied in the dynamical system of two degrees of freedom (2D), while for the dynamical system of three degrees of freedom (3D), described by the Hamiltonian (6), will shall use the $S(w)$ spectrum. The layout of this article is as follows: In Section 2, we present an analysis of the phase plane of the 2D Hamiltonian system. Furthermore, we apply the $S(g)$ dynamical spectrum in order to study the various resonant orbits and the different chaotic components. In the same Section, the evolution of the 2D sticky orbits is investigated using the $S(g)$ dynamical spectrum. In Section 3, we investigate the properties of the 3D Hamiltonian system, using the new $S(w)$ dynamical spectrum. Once more, we test and prove the efficiency of our new indicator by studying the 3D resonant orbits, the different chaotic components and also the sticky regions. Finally, we close with Section 4 in which we present a discussion and the conclusions of the present research.

\section{Application of the $S(g)$ dynamical spectrum in the 2D Hamiltonian system}

In this Section, we shall study the nature of orbits in the 2D dynamical model, using the $S(g)$ spectrum. The corresponding Hamiltonian for the 2D system $\left(z=p_z=0\right)$ is
\begin{equation}
H_2 = \frac{1}{2} \left(p_x^2 + p_y^2 \right) + V_{tot}(x,y) = h_2,
\end{equation}
where $h_2$ is the numerical value of 2D Hamiltonian.
\begin{figure}[!tH]
\centering
\resizebox{\hsize}{!}{\rotatebox{0}{\includegraphics*{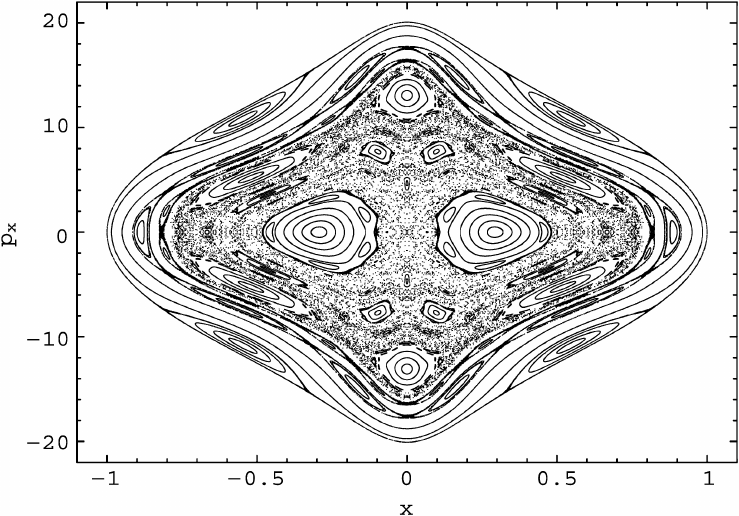}}}
\caption{The $\left(x, p_x\right)$, $y=0$, $p_y >0$ phase plane, when $h_2=1.5$. Details are given in the text.}
\end{figure}

Figure 1 shows the $\left(x,p_x\right)$, $y=0$, $p_y>0$ Poincar\'{e} phase plane for the Hamiltonian (8) when $h_2=1.5$. The values of the involved parameters are: $\omega =10, a=4, M_n=50$ and $c_n=0.25$. We observe, that the phase plane shown in Fig. 1 is very interesting and complicated. There is a plethora of sets of islands of invariant curves which correspond to resonant orbits of lower and higher multiplicity. Moreover, we discern the presence of several different chaotic components and also considerable sticky regions, embedded in the unified chaotic domain. These interesting features of the phase plane will help us to test, verify and eventually prove the efficiency and the reliability of the new dynamical spectrum.
\begin{figure*}
\centering
\resizebox{\hsize}{!}{\rotatebox{0}{\includegraphics*{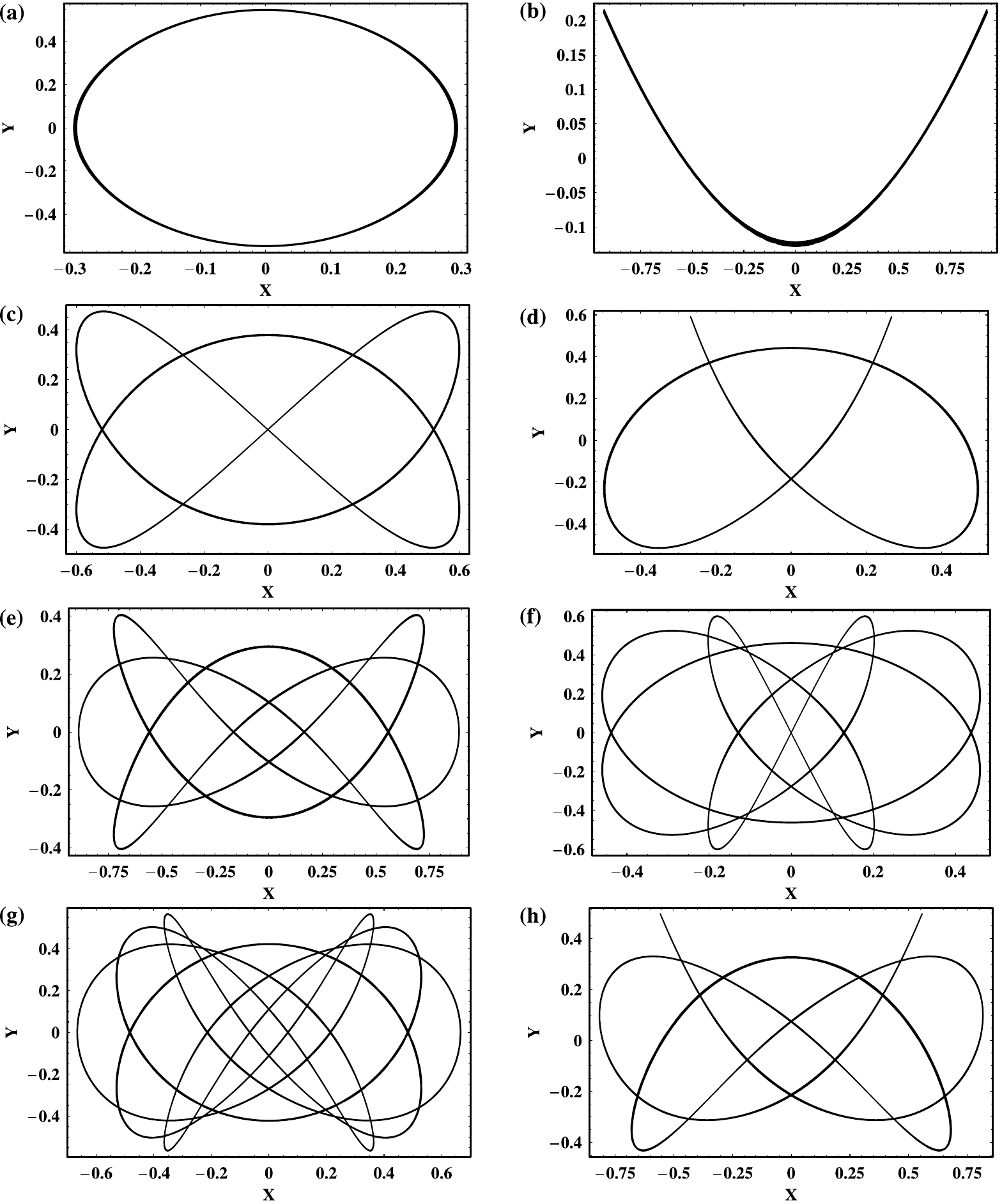}}}
\vskip 0.01cm
\caption{(a-h): Different types of resonant periodic orbits in the 2D dynamical system. The initial conditions and the values of all the involved parameters are given in the text.}
\end{figure*}

It is therefore evident from Fig. 1, that this dynamical system presents a large variety of resonant orbits, which correspond to sets of multiple islands of invariant curves in the $\left(x,p_x\right)$ phase plane. Figure 2a-h shows eight representative periodic orbits for the 2D system. The values of all the other parameters are as in Fig. 1. Figure 2a shows a periodic orbit characteristic of the 1:1 resonance with initial conditions $x_0=0.2943$ and $p_{x0}=0$. For all 2D orbits $y_0=0$, while the value of $p_{y0}$ is always found from the energy integral (8). In Figure 2b one may observe a 1:2 periodic orbit with initial conditions $x_0=0.548$ and $p_{x0}=10.8$. Figure 2c depicts a 2:3 resonant periodic orbit with initial conditions $x_0=0$ and $p_{x0}=13.07$. Moreover, in Figure 2d we can see a periodic orbit of the 3:4 resonance. The initial conditions are $x_0=0.097$ and $p_{x0}=7.7$. Figure 2e presents a 3:5 resonant periodic orbit with initial conditions $x_0=0.889$ and $p_{x0}=0$. Similarly, in Figure 2f a 4:5 resonant periodic orbit is presented. The initial conditions are $x_0=0$ and $p_{x0}=4.675$. In Figure 2g we observe a complicated resonant periodic orbit with initial conditions $x_0=0.667$ and $p_{x0}=0$, which belongs to the family of the 5:7 resonance. Finally, a periodic orbit of the 5:8 resonance is presented in Figure 2h. The initial conditions are $x_0=0.803$ and $p_{x0}=2.03$. The integration time for all orbits shown in Fig. 2a-h is 100 time units. It is remarkable, that all the above resonant families of orbits exist simultaneously in this 2D Hamiltonian system. The interaction of all these resonances, justifies the presence of the different sticky regions and also the different chaotic components displayed in the phase plane of Fig. 1.
\begin{figure*}
\centering
\resizebox{\hsize}{!}{\rotatebox{0}{\includegraphics*{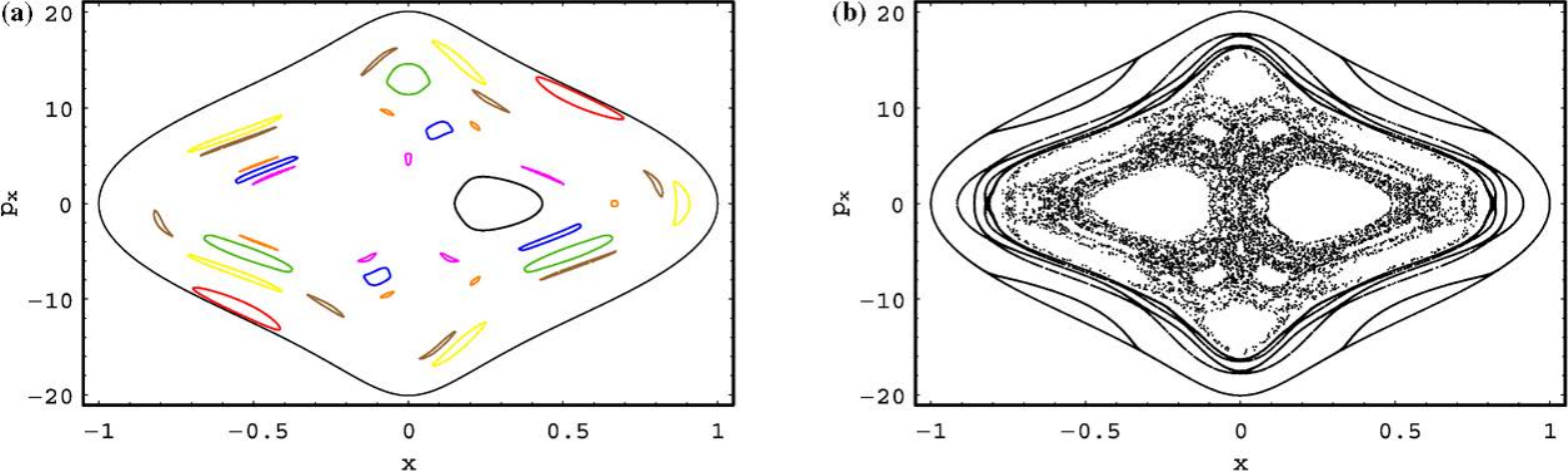}}}
\vskip 0.01cm
\caption{(a-b): (a-\textit{left}): The most significant sets of islands of invariant curves in the $\left(x,p_x\right)$ phase plane and (b-\textit{right}): the three different chaotic components together with the unified chaotic sea. (Color figurer online)}
\end{figure*}

In Figure 3a we present with different colors the most significant sets of islands of invariant curves in the $\left(x,p_x\right)$ phase plane. These islands of invariant curves correspond to different types of families of resonant orbits. The large central island is marked in black and corresponds to the 1:1 resonance. The initial conditions are $x_0=0.15$ and $p_{x0}=0$. The islands of invariant curves corresponding to the 1:2 resonance are red and the initial conditions are $x_0=0.5$ and $p_{x0}=12.25$. The islands of invariant curves which correspond to the 2:3 resonance are colored in green. The initial conditions are $x_0=0$ and $p_{x0}=14.6$. Moreover, the 3:4 resonant islands with initial conditions $x_0=0.1$ and $p_{x0}=8.5$ are plotted in blue. The set of islands of invariant curves which corresponds to the 3:5 resonance is marked in yellow. The initial conditions are $x_0=0.909$ and $p_{x0}=0$. The islands of invariant curves produced by the 4:5 resonance are depicted with magenta color and the initial conditions are $x_0=0$ and $p_{x0}=4.021$. The set of islands of invariant curves corresponding to 5:7 resonance is orange. The initial conditions are $x_0=0.657$ and $p_{x0}=0$. Last but not least, the islands of invariant curves produced by the 5:8 resonance are plotted in brown and the initial conditions are $x_0=0.802$ and $p_{x0}=1.44$. The outermost black solid line shown in Fig. 3a is the Zero Velocity Curve (ZVC). The integration time for all the sets of invariant curves shown in Fig. 3a is 2000 time units. The three different chaotic components together with the unified chaotic sea are shown in Figure 3b. The first chaotic component (CC-1) has initial conditions $x_0=-0.82$ and $p_{x0}=0$, the second (CC-2) $x_0=-0.86$ and $p_{x0}=0$, while the initial conditions for the third chaotic component (CC-3) are $x_0=-1$ and $p_{x0}=0$. Here, we notice that the third chaotic component coincides with the ZVC. All the chaotic components and the chaotic sea presented in Fig. 3b were integrated for 2000 time units. The equation of the limiting curve - ZVC (that it the curve containing all the invariant curves of the 2D system) is
\begin{equation}
\frac{1}{2}p_x^2 + V_{tot}(x) = h_2.
\end{equation}

In an earlier work [39] (hereafter Paper II) we introduced and used a new dynamical parameter the $S(g)$ spectrum, in order to study the islandic motion of resonant orbits and the evolution of sticky regions. The $S(g)$ spectrum was applied in a axially symmetric galactic model in the meridian $\left(r,z\right)$ plane. However, it is easy enough to modify its definition in order to make the $S(g)$ spectrum applicable in Cartesian coordinates $(x,y)$. Thus, we define the dynamical parameter $g$ as
\begin{equation}
g_i = \frac{x_i + p_{xi} - x_i p_{xi}}{p_{yi}},
\end{equation}
where $\left(x_i, p_{xi}, p_{yi}\right)$ are the successive values of the $(x, p_x, p_y)$ elements of the 2D orbit on the Poincar\'{e} $\left(x,p_x\right)$, $y=0$, $p_y>0$ phase plane. We shall define the dynamical spectrum of the parameter $g$, through its distribution function
\begin{equation}
S(g) = \frac{\Delta N(g)}{N \Delta g},
\end{equation}
where $\Delta N(g)$ are the number of the parameters $g$ in the interval $\left(g, g + \Delta g\right)$, after $N$ iterations. By definition, the $g$ parameter is based on a combination of the coordinates and the velocities of the 2D orbit. It is very interesting to test the effectiveness of the $S(g)$ spectrum in the present dynamical system and find out if it has the ability to identify all the different types of resonant orbits. Moreover, it would be extremely beneficial, if it could distinguish the different chaotic components and also if it is sensitive enough in order to provide reliable results regarding the evolution of the 2D sticky orbits.
\begin{figure*}
\centering
\resizebox{\hsize}{!}{\rotatebox{0}{\includegraphics*{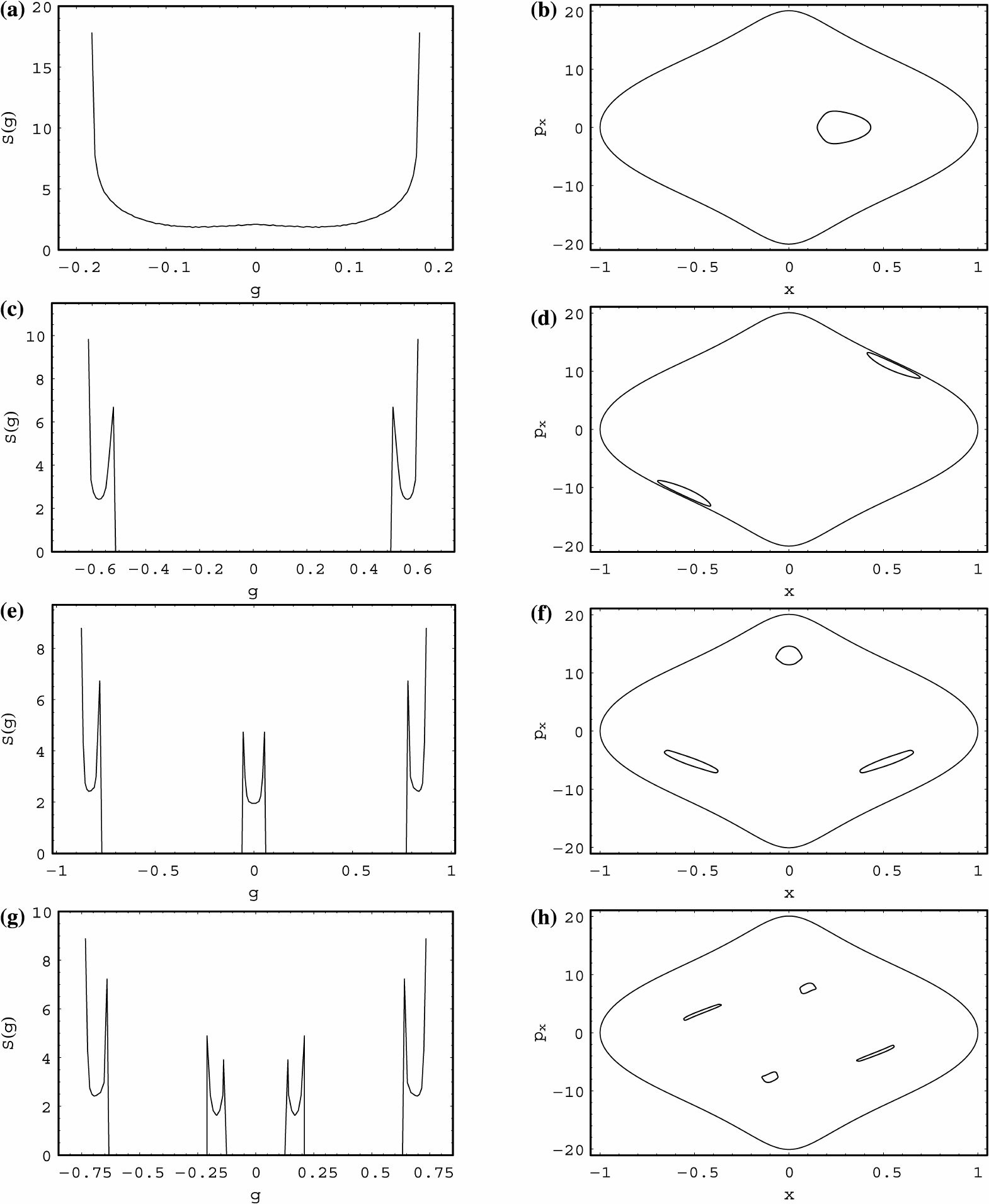}}}
\vskip 0.01cm
\caption{(a-h): (\textit{left pattern}): The $S(g)$ dynamical spectrum for the different types of resonant orbits and (\textit{right pattern}) the corresponding sets of islands of invariant curves in the $\left(x, p_x\right)$ phase plane.}
\end{figure*}
\begin{figure*}
\centering
\resizebox{\hsize}{!}{\rotatebox{0}{\includegraphics*{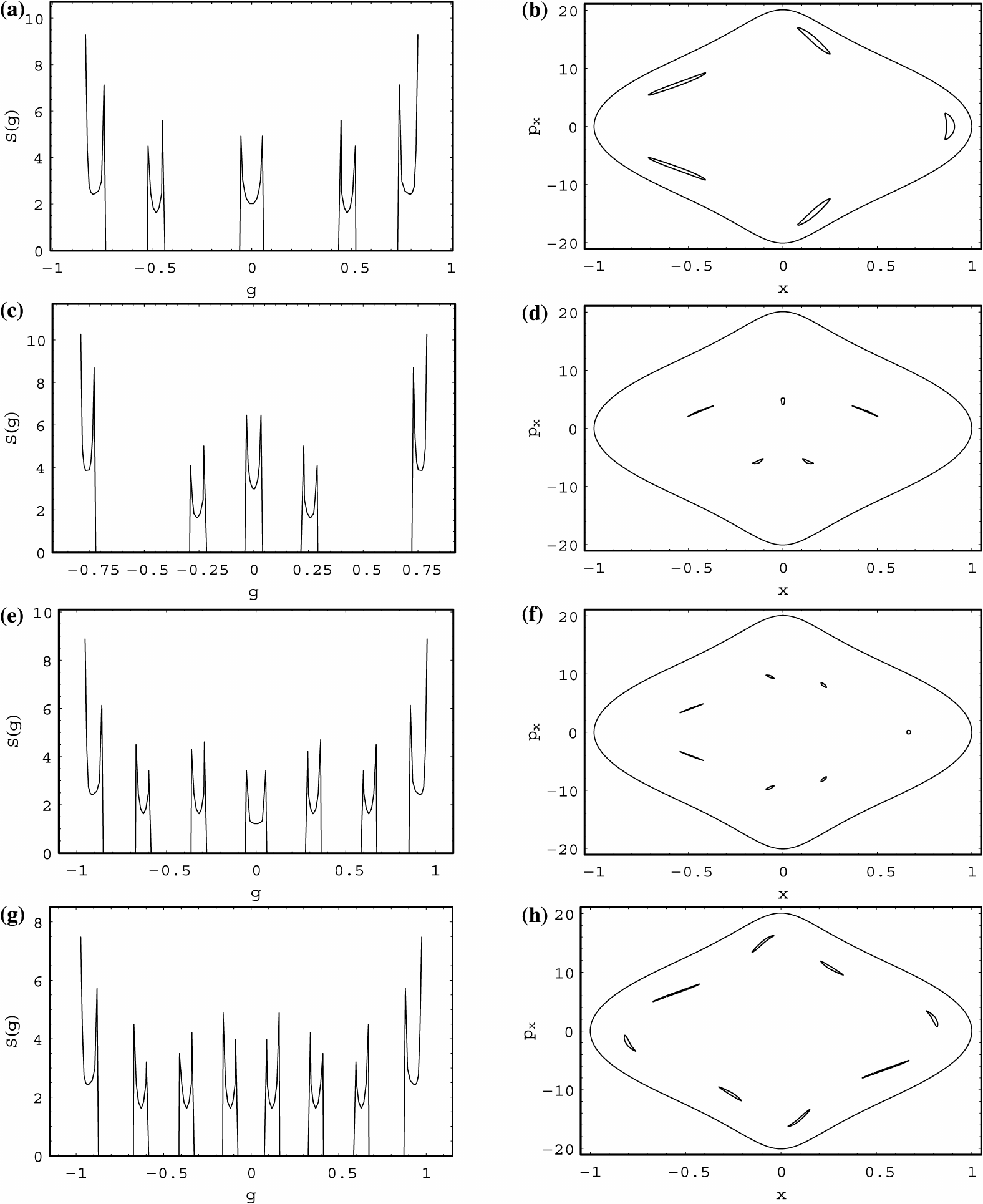}}}
\vskip 0.01cm
\caption{(a-h): (\textit{left pattern}): The $S(g)$ dynamical spectrum for the different types of resonant orbits and (\textit{right pattern}) the corresponding sets of islands of invariant curves in the $\left(x, p_x\right)$ phase plane.}
\end{figure*}

First we shall apply the $S(g)$ dynamical spectrum, in order to identify the different types of resonant orbits which correspond to the sets of islands of invariant curves shown in Fig. 3a. The left side of Figure 4a-h shows the $S(g)$ spectrums for the resonant orbits, while the right side of the same figure presents the corresponding islands of invariant curves in the $\left(x, p_x\right)$ phase plane. Figure 4a shows the $S(g)$ spectrum for the 1:1 resonant orbit with initial conditions as in Fig. 3a. As expected, one can observe a well defined ``$U$-shaped" structure with two abrupt peaks in the two extreme opposite values of $g$, indicating regular motion. The corresponding island is shown in Figure 4b. ``$U$-shaped" spectra are characteristic of quasi-periodic orbits with initial conditions very close to the stable fixed points [37]. The existence of these abrupt peaks can be explained theoretically. In fact such peaks are infinities [13,28]. These infinities have to be distinguished from local smooth maxima, which may be present in some spectra. The outermost black solid line shown in Fig. 4b is the ZVC. Figure 4c depicts the $S(g)$ spectrum for the 1:2 resonant orbit with initial conditions as in Fig. 3a, while in Figure 4d we see the two islands of invariant curves in the $\left(x, p_x\right)$ phase plane. In Figure 4e we present the $S(g)$ spectrum for the 2:3 resonant orbit of Fig. 3a. In this case, we observe three well defined $U$ type structures, as much is the number of islands of invariant curves shown in Figure 4f. Moreover, looking more carefully we see that the left and the right spectra are exactly symmetrical about the $g=0$ axis, while the central spectrum lies on both sides of this axis. This indicates that two of the islands of invariant curves are symmetrical about the $p_x$ axis, while the third intersects the $p_x$ axis. In other words, we have the case of a quasi-periodic orbit, with a starting point on the $p_x$ axis. Figure 4g and 4h are similar to Fig. 4e and 4f but for the 3:4 resonant orbit of Fig. 3a producing four islands of invariant curves.

Figure 5a illustrates the $S(g)$ spectrum for the 3:5 resonant orbit of Fig. 3a, while in Figure 5b the corresponding islands of invariant curves are presented. Similarly, in Figure 5c we see the $S(g)$ spectrum for the 4:5 resonant orbit of Fig. 3a and in Figure 5d the five islands of invariant curves in the $\left(x, p_x\right)$ phase plane. Furthermore, Figure 5e depicts the $S(g)$ dynamical spectrum for the 5:7 resonant orbit of Fig. 3a, while in Figure 5f we see the set of seven small islands of invariant curves. Finally, in Figure 5g one may observe the structure of the $S(g)$ spectrum for the 5:8 resonant orbit with initial conditions as in Fig. 3a. The corresponding set of the eight small islands is given in Figure 5h. For all the $S(g)$ spectrums presented in Figs. 4 an 5 the initial conditions are as in Fig. 3a, while the integration time equal to 5000 time units. From the above results, we may conclude that the $S(g)$ spectrum has the ability to identify every type of resonant orbit, since it produces always as many $U$ type structures as much as the total number of islands of invariant curves in the $\left(x, p_x\right)$ phase plane. Moreover, it provides fast and very reliable results not only for simple resonant 2D orbits, but also for complicated resonant orbits of higher multiplicity which correspond to sets of multiple islands of invariant curves.
\begin{figure*}
\centering
\resizebox{\hsize}{!}{\rotatebox{0}{\includegraphics*{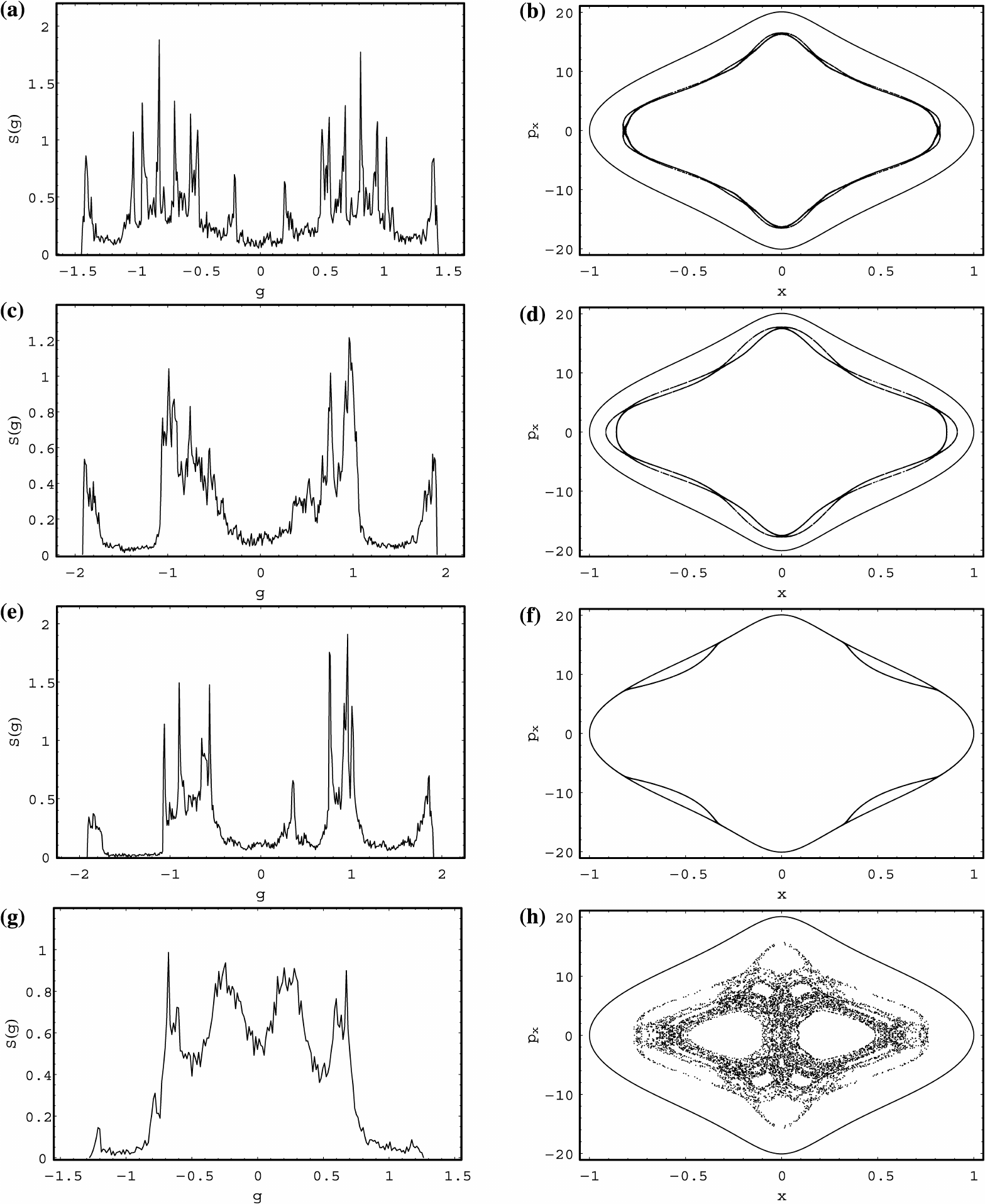}}}
\vskip 0.01cm
\caption{(a-h): (\textit{left pattern}): The $S(g)$ dynamical spectrum for the different chaotic components and the unified chaotic sea and (\textit{right pattern}) the corresponding chaotic components and the unified chaotic sea in the $\left(x, p_x\right)$ phase plane.}
\end{figure*}

Our next step, is to deploy the $S(g)$ spectrum in order to identify the several chaotic components of the 2D dynamical system. Figure 6a shows the $S(g)$ spectrum for the first chaotic component (CC-1). We observe a complicated, highly asymmetric structure, with a lot of large, small and abrupt peaks, indicating chaotic motion. The first chaotic component (CC-1) inside the ZVC is shown in Figure 6b. Similarly, Figure 6c depicts the $S(g)$ spectrum for the second chaotic component (CC-2), while CC-2 is presented in Figure 6d. Moreover, Figure 6e depicts the $S(g)$ dynamical spectrum for the third chaotic component (CC-3), while CC-3 is presented in Figure 6f. In Figure 6g one may observe the $S(g)$ spectrum for the unified chaotic sea which is shown in Figure 6h. Here, we must point out that Figs. 6a, 6c, 6e and 6g reveal the sensitivity of the $S(g)$ spectrum. It is obvious that the structure shown in Fig. 6g is quite different from those shown in Figs. 6a, 6c and 6e. In other words, the $S(g)$ spectrum is sensitive enough allowing us to know whether an orbit corresponds to a chaotic sea or a chaotic component. The initial conditions for the $S(g)$ spectra corresponding to the chaotic components and the chaotic sea, shown in Fig. 6 are as in Fig. 3b, while the integration time is equal to 5000 time units.
\begin{figure*}
\centering
\resizebox{\hsize}{!}{\rotatebox{0}{\includegraphics*{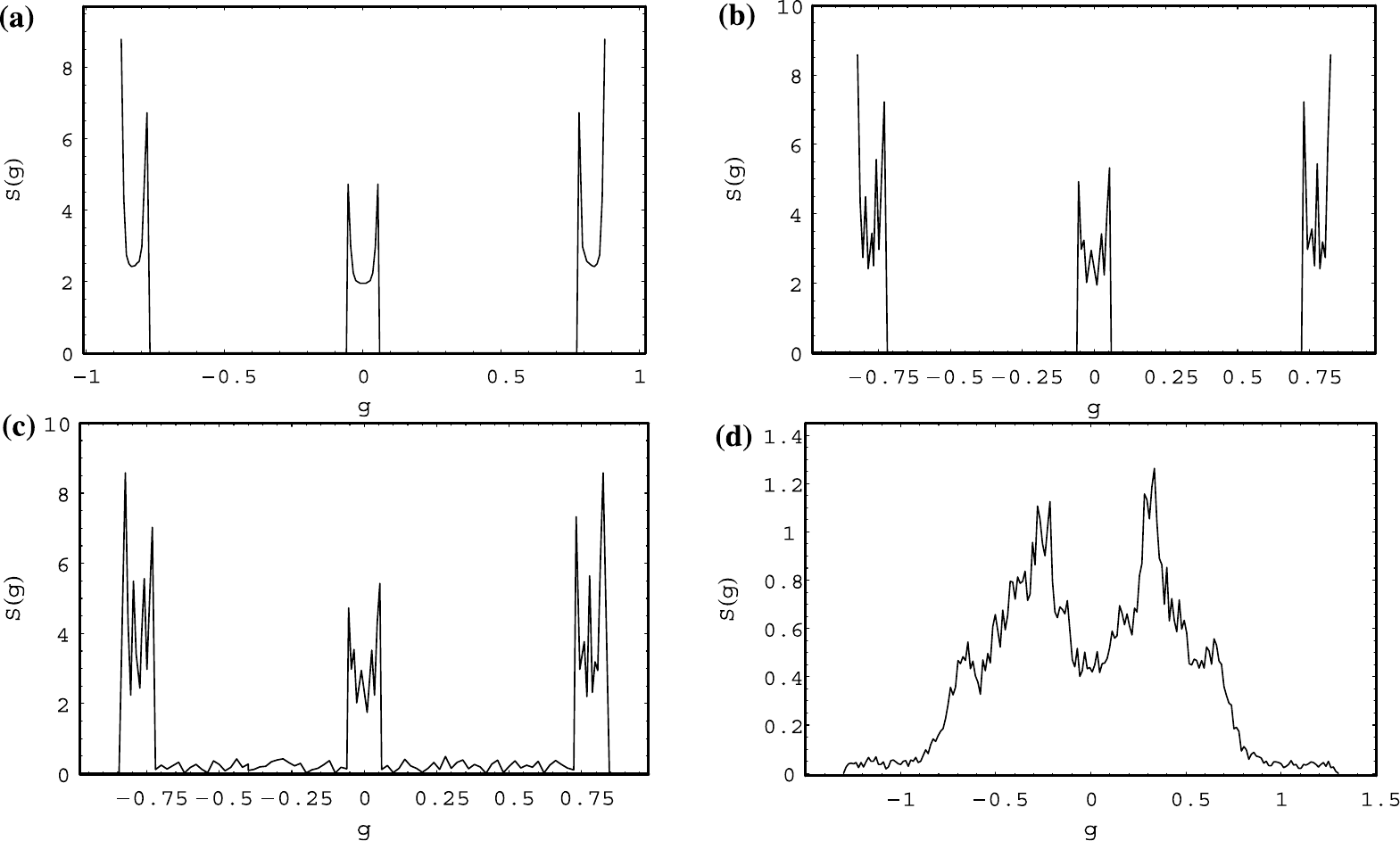}}}
\vskip 0.01cm
\caption{(a-d): (a-\textit{upper left}): The $S(g)$ dynamical spectrum of a 2:3 resonant quasi-periodic 2D orbit and (b-d): the time evolution of the $S(g)$ spectrum for a nearby sticky orbit. See the text for details.}
\end{figure*}
\begin{figure*}
\centering
\resizebox{\hsize}{!}{\rotatebox{0}{\includegraphics*{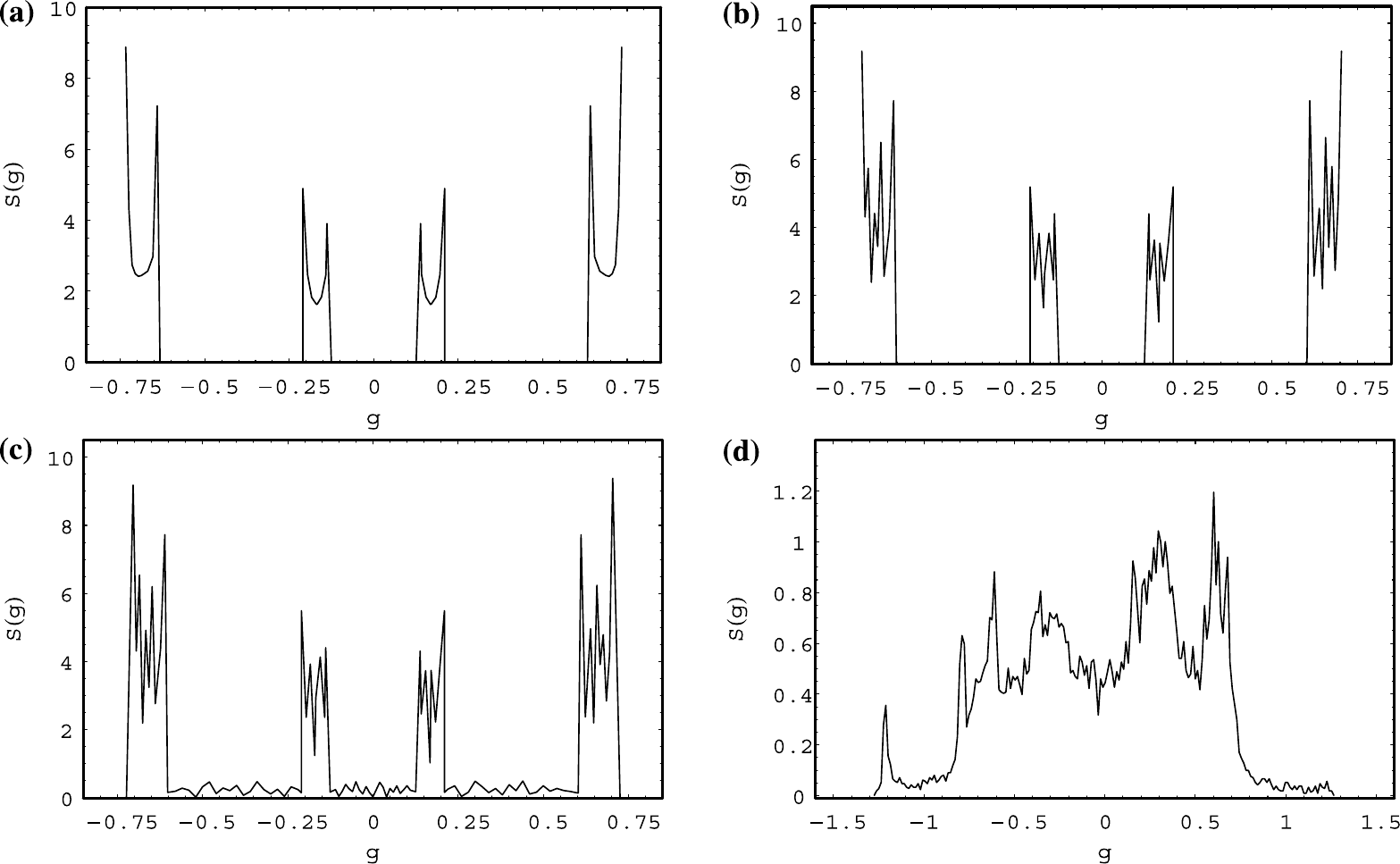}}}
\vskip 0.01cm
\caption{(a-d): (a-\textit{upper left}): The $S(g)$ dynamical spectrum of a 3:4 resonant quasi-periodic 2D orbit and (b-d): the time evolution of the $S(g)$ spectrum for a nearby sticky orbit. See the text for details.}
\end{figure*}

Our final step, is to verify if the $S(g)$ spectrum can provide reliable results regarding the evolution of 2D sticky orbits. Figure 7a shows the $S(g)$ spectrum of a 2:3 resonant 2D orbit producing three islands of invariant curves. The initial conditions are $x_0=0, y_0=0, p_{x0}=14.6$, while the value of $p_{y0}$ is always found from the energy integral (8). The values of all other parameters are as in Fig. 1. Here, we observe three well defined $U$ type spectra, indicating regular motion. Note, that the number of spectra is equal to the number of islands. Figure 7b shows the $S(g)$ spectrum of an orbit starting near the above regular orbit. The initial conditions are $x_0=0, y_0=0, p_{x0}=10.5$. Here, we observe three different spectra with a lot of large and small peaks. This indicates that we have a sticky orbit and the sticky region is composed of three sticky islands. The sticky period is about $T=740$ time units. Figure 7c shows the spectrum of the same orbit, when $T=790$ time units. Here, the three spectra have joined in order to produce a unified structure. This indicates, that after the sticky period the test particle has gone to the chaotic sea. In Figure 7d we can see the $S(g)$ spectrum of the orbit when $T=5000$ time units. Here, we observe a spectrum of a highly chaotic orbit with a lot of small, large, asymmetric and abrupt peaks. Note that, the sticky period obtained from the $S(g)$ spectrum is very close to the sticky period obtained by the $S(c)$ spectrum in Paper II (see for more details in Fig. 3 in Paper II).

Let us now follow the time evolution of another 2D sticky orbit using the $S(g)$ spectrum. Figure 8a shows the $S(g)$ spectrum of a 3:4 resonant 2D orbit producing four islands of invariant curves. The initial conditions are $x_0=0.1, y_0=0, p_{x0}=8.5$, while the value of $p_{y0}$ is always found always from the energy integral (8). The values of all other parameters are as in Fig. 1. Here, we observe four well defined $U$ type spectra, indicating regular motion. Note, that the number of spectra is once more equal to the total number of islands. Figure 8b shows the $S(g)$ spectrum of an orbit starting near the above regular orbit. The initial conditions are $x_0=0.1, y_0=0, p_{x0}=7.45$. Here, we observe four different spectra with a lot of large and small peaks. This indicates that we have a sticky orbit and the sticky region is composed of four sticky islands. The sticky period is about $T=950$ time units. Figure 8c shows the spectrum of the same orbit, when $T=1000$ time units. Here, the four spectra have joined in order to produce a unified structure. This indicates, that after the sticky period the test particle has left completely from the sticky region and moved to the chaotic sea. In Figure 8d we can see the $S(g)$ spectrum of the same orbit when $T=5000$ time units. Here, we observe a highly chaotic spectrum with a lot of small, large, asymmetric and abrupt peaks. Therefore, it is evident that the $S(g)$ spectrum can be deployed in order to calculate the sticky period of a 2D sticky orbit and also to follow its time evolution towards the chaotic sea.
\begin{figure}[!tH]
\centering
\resizebox{\hsize}{!}{\rotatebox{0}{\includegraphics*{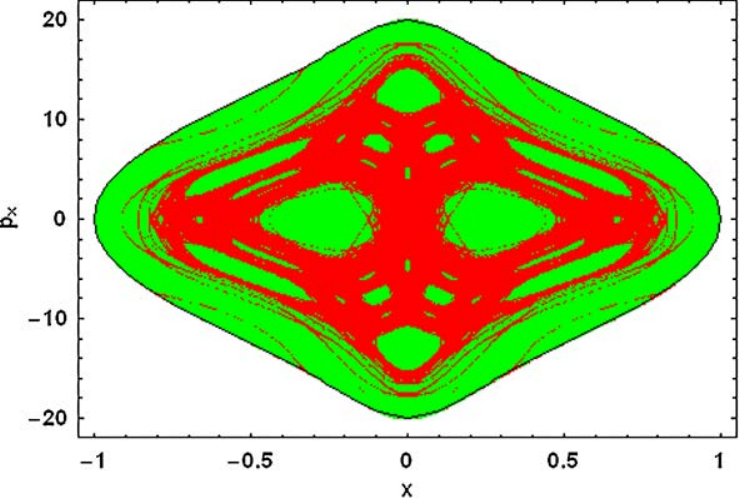}}}
\caption{The structure of the $(x, p_x)$ phase plane reproduced using the $S(g)$ spectrum. Inside the \textit{green regions} the motion is ordered, while inside the \textit{red regions} the motion is chaotic. The outermost \textit{black solid line} is the Zero Velocity Curve (ZVC). (Color figure online)}
\end{figure}

So far, we have seen that the $S(g)$ dynamical spectrum can provide very reliable results when we study the character of the orbits in a 2D Hamiltonian system. It has the ability to identify complicated resonant orbits, different chaotic components and also the evolution of sticky orbits. However, all the above results obtained from the $S(g)$ are qualitative. Therefore, we must ispect each time the shape of the spectrum's structure by eye, in order to characterize a 2D orbit. Obviously, this is not very practical when someone wants to check a large number of orbits, so as to form an idea about the global structure of the phase plane of a dynamical system. Thus, we need to establish a criterion, in order to quantify the results given by the $S(g)$ spectrum. This criterion can be derived by looking the structure of the spectra shown in Figs. 4, 5, 6, 7 and 8. One may observe, that when the orbit is regular the maximum value of $S(g)$, that is $S(g)_{max}$, is high $\left(S(g)_{max} \geq 7 \right)$. On the contrary, when the orbit is chaotic $S(g)_{max}$ is significantly smaller $\left(S(g)_{max} \leq 3 \right)$. By testing a large number of orbits (approximately 1000), we conclude that when $S(g)_{max} \geq 2.5$ the orbit is regular, while when $S(g)_{max} < 2.5$ the orbit is chaotic. By using the above criterion for initial conditions $(x_0,p_{x0})$ on the entire phase plane of Fig. 1 and giving to each initial condition a color according to the value of the $S(g)_{max}$, we can have a clear picture of the regions where chaotic or ordered motion occurs. The outcome of this procedure for the 2D dynamical system (8), using a dense grid of initial conditions on the phase plane, is presented in Figure 9. Thus, in Fig. 9 we clearly distinguish between green regions, where the motion is ordered and red regions, where it is chaotic. The outermost black solid line is the Zero Velocity Curve (ZVC). It is worth mentioning that in Fig. 9 we can observe small islands of stability inside the large unified chaotic sea, which are not visible in the detailed phase plane of Fig. 1. Although each orbit in Fig. 9 was computed for only 1000 time units, this time interval was sufficient for the clear revelation of small ordered regions inside the chaotic sea. For a grid of about 6 $\times$ $10^4$ equally spaced initial conditions $(x_0,p_{x0})$, we need about 7 hours of CPU time on a Pentium Dual-Core 2.2GHz PC, in order to construct the grid presented in Fig. 9.

\section{Application of the $S(w)$ dynamical spectrum in the 3D Hamiltonian system}

In this Section, we shall present results regarding the character of motion in the 3D dynamical system. For this purpose we use the 3D Hamiltonian (6) and take $h=h_2$, that is the value of energy is equal to that of the 2D system. Now, we can use the results obtained from the 2D system, in order to investigate the motion in the 3D system. What we do is to consider orbits with a starting a point on the $\left(x,p_x\right)$ phase plane with an additional value of $z=z_0$ and follow the evolution of the 3D orbits.
\begin{figure*}
\centering
\resizebox{0.75\hsize}{!}{\rotatebox{0}{\includegraphics*{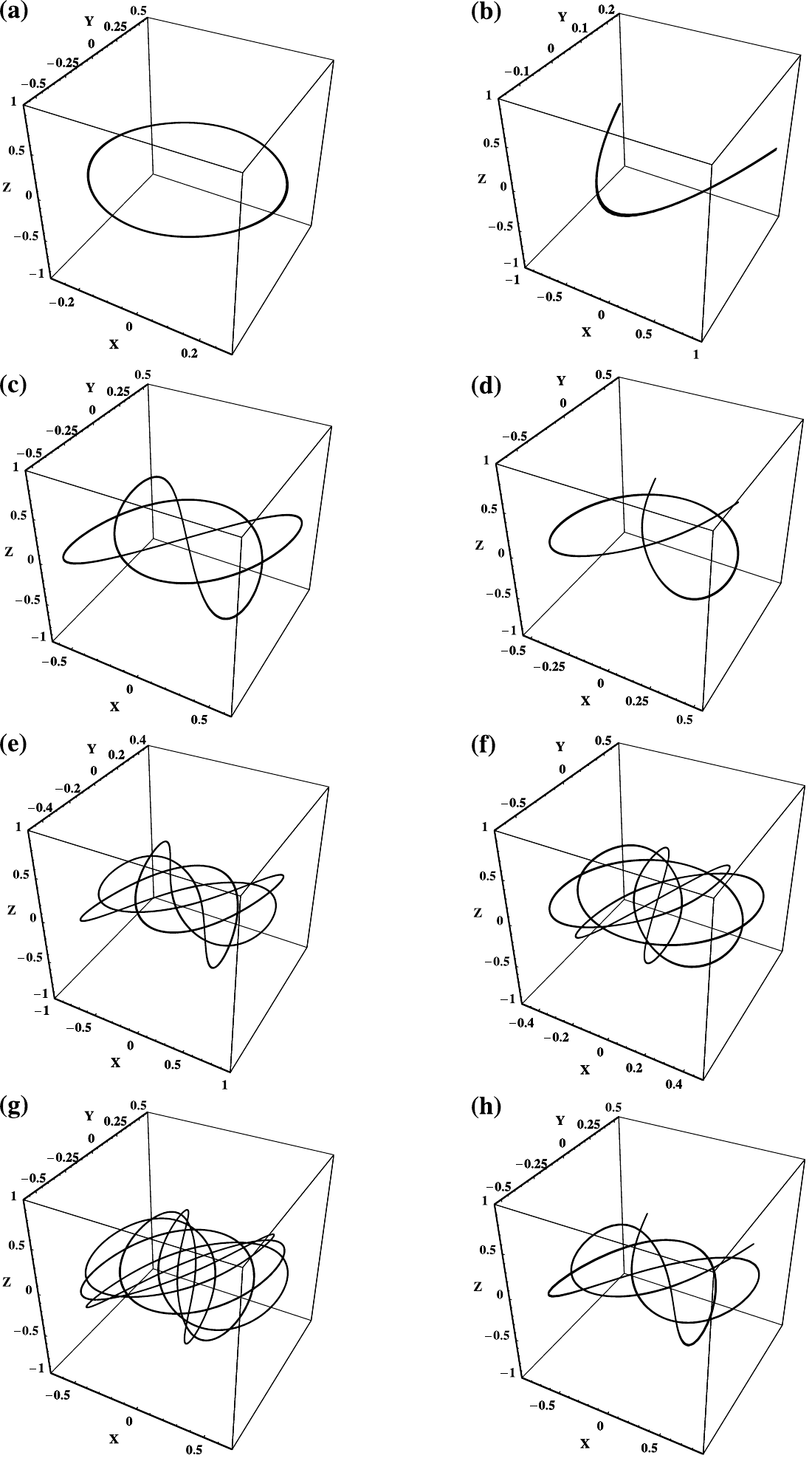}}}
\vskip 0.01cm
\caption{(a-h): Typical resonant periodic orbits in the 3D system. The initial conditions for all the orbits and additional details are provided in the text.}
\end{figure*}

Figure 10a-h presents eight typical resonant periodic orbits in the 3D system. In all resonant orbits the values of the initial conditions $(x_0, p_{x0})$ are the same as in the corresponding 2D orbits, discussed in Fig. 2a-h. For all 3D orbits $y_0=p_{z0}=0$, the value of $p_{y0}$ is always found from the energy integral (6), while the initial value of $z_0$ is equal to 0.005. Note, that all orbits shown in Fig. 10a-h stay very close to the galactic plane, due to the small initial value of $z_0$. Nevertheless, we observe that even for a small value of $z_0$ the large variety of resonant orbits we studied in the 2D system in the previous Section, still exists also in the 3D dynamical system. The integration time for all 3D orbits shown in Fig. 10a-h is 100 time units.

A basic problem in Hamiltonian systems of three degrees of freedom is the visualization of the 4D surfaces of section. Let us assume the phase space of an autonomous Hamiltonian system, that has 6 dimensions, e.g. in Cartesian coordinates, $\left(x, y, z, \dot{x}, \dot{y}, \dot{z}\right)$. For a given value of the Hamiltonian a trajectory lies on a 5D manifold. In this manifold the surface of section is 4D. This does not allow us to visualize it directly. Several methods have been applied to overcome this problem in the past and we summarize them below.

The structure of the 4D space phase space was examined for the first time in the pioneer work of Froeschl\'{e} [14,15]. In that work, he used stereoscopic views and the method of slices in order to understand and interpret the structure of the tori, that appeared at the neighborhood of stable periodic orbits. Similar methods have also been applied for studying the 3D projections of invariant tori in the 4D surface of section or in the phase space of a 4D symplectic map [8,24,25]. The 2D projections of such invariant tori have been examined on various 2D planes in detail [33,34]. In the present work, we shall use the method proposed by Patsis \& Zachilas [27]. In this method, we first define the surface of section that we will use, e.g. $y = 0$ with $p_y > 0$. Then, we select a 3D subspace projection e.g. $\left(x, p_x, z\right)$ of the $\left(x, p_x, z, p_z\right)$ 4D phase space. If the 3D projection is a well formed 3D invariant torus then the initial conditions correspond to a regular orbit. On the other hand, if instead of a well defined torus, we observe a vague cloudy structure in the $\left(x, p_x, z\right)$ projection, then the corresponding 3D orbit is chaotic.

Another interesting approach is the method introduced by Pfenniger [29] This method has been used by several researchers in the past [4,32]. We take sections in the plane $y = 0$, $p_y > 0$ of 3D orbits, whose initial conditions differ from the plane parent periodic orbits only by the $z$ component. The set of the resulting 4-dimensional points in $\left(x, p_x, z, p_z\right)$ phase space are projected on the $(z, p_z)$ plane. If the projected points lie on a well defined curve, lets call it an ``invariant curve", then the motion is regular, while if not, the motion is chaotic. The projected points on the $(z, p_z)$ plane show nearly invariant curves around the periodic points at $z = 0$, $p_z = 0$, as long as the coupling is weak. When the coupling is stronger the corresponding projections in $(z, p_z)$ plane display increasing departure of the plane periodic point, up to making a direct orbit a retrograde one and vise versa. Here, we must define what one means by direct and retrograde 3D orbit. If consequents in the $(z, p_z)$ section of the 3D orbit drop in one of the two domains of the corresponding section of 2D orbits at the same value of the Jacobi integral $E_J$, we can distinguish between direct and retrograde motion. Orbits which visit both domains are intermittently direct or retrograde.

For the study of 3D orbits, we introduced and used by Zotos [40] (hereafter Paper III) the $S(w)$ spectrum. Here we must remind to the reader that the $S(w)$ spectrum is the distribution function of the $w$ parameter
\begin{equation}
S(w) = \frac{\Delta N(w)}{N \Delta w},
\end{equation}
where $\Delta N(w)$ are the number of the parameters $w$ in the interval $\left(w, w+\Delta w \right)$, after $N$ iterations. The parameter $w$, is defined as
\begin{equation}
w_i = \frac{(x_i - p_{x_i})-(z_i - p_{zi})}{p_{yi}},
\end{equation}
where $\left(x_i, z_i, p_{xi}, p_{yi}, p_{zi}\right)$ are the successive values of the $(x, z, p_x, p_y, p_z)$ elements of the 3D orbit. By definition, the $w$ parameter is based on a combination of coordinates and velocities of the 3D orbit. It would be very interesting to test the effectiveness of the $S(w)$ spectrum in the present dynamical system and find out if it has the ability to identify all the 3D resonant orbits. Moreover, it would be extremely beneficial, if it could distinguish the different chaotic components and also if it is sensitive enough in order to provide reliable results regarding the evolution of the 3D sticky orbits.
\begin{figure*}
\centering
\resizebox{0.75\hsize}{!}{\rotatebox{0}{\includegraphics*{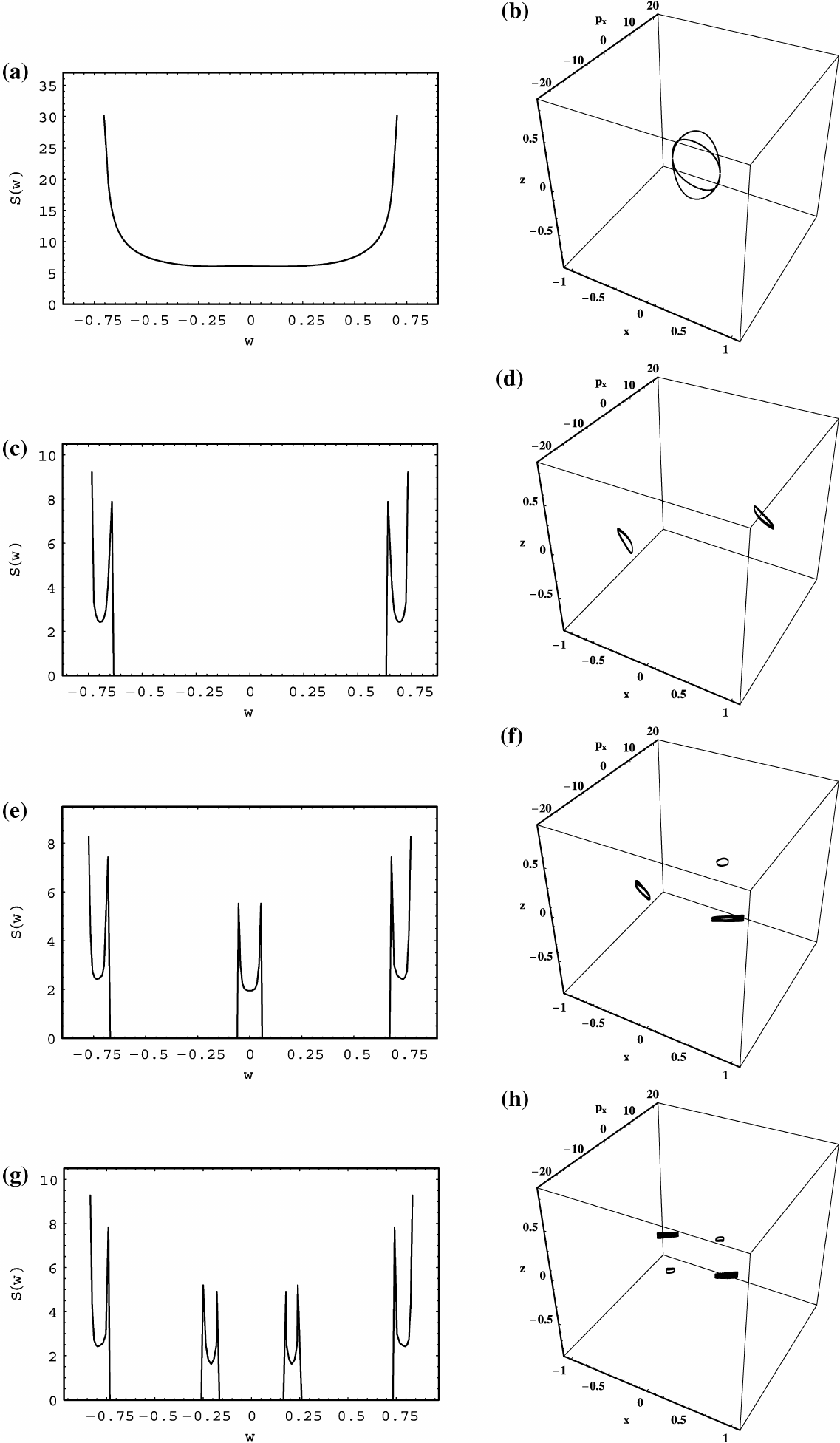}}}
\vskip 0.01cm
\caption{(a-h): (\textit{left pattern}): The $S(w)$ dynamical spectrum for different types of resonant 3D orbits and (\textit{right pattern}) the corresponding 3D islands of invariant tori in the $\left(x, p_x, z\right)$ phase subspace.}
\end{figure*}
\begin{figure*}
\centering
\resizebox{0.75\hsize}{!}{\rotatebox{0}{\includegraphics*{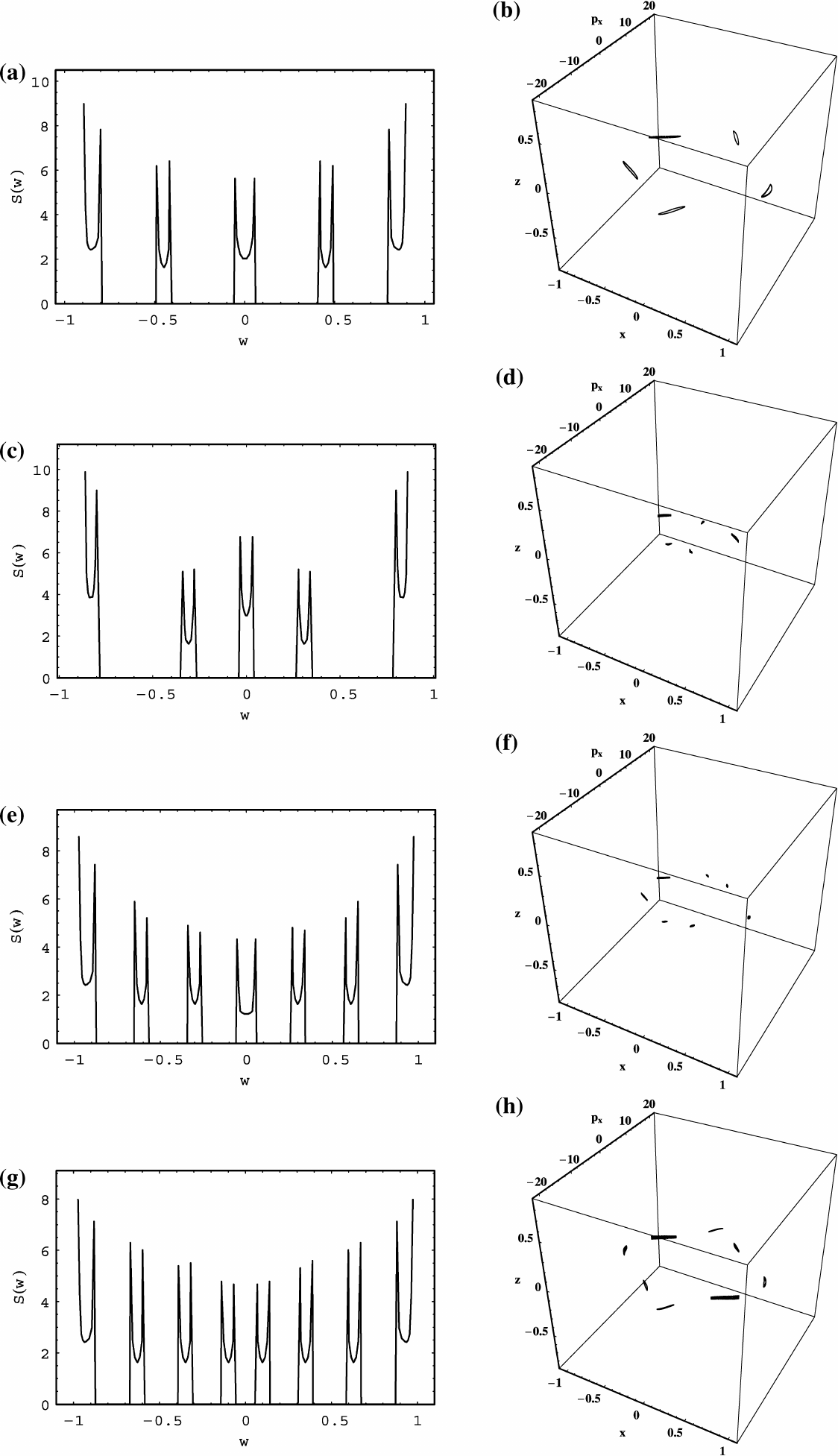}}}
\vskip 0.01cm
\caption{(a-h): (\textit{left pattern}): The $S(w)$ dynamical spectrum for different types of resonant 3D orbits and (\textit{right pattern}) the corresponding 3D islands of invariant tori in the $\left(x, p_x, z\right)$ phase subspace.}
\end{figure*}

We shall first apply the $S(w)$ dynamical spectrum, in order to identify the different types of 3D resonant orbits. The left side of Figure 11a-h shows the $S(w)$ spectrums for the 3D resonant orbits, while the right side of the same figure presents the corresponding 3D invariant tori in the $\left(x, p_x, z\right)$ phase subspace. Figure 11a shows the $S(w)$ spectrum for the 1:1 resonant 3D orbit with initial conditions $\left(x_0, p_{x0}\right)$ as in Fig. 3a. As expected, one can observe a well defined $U$ type structure, indicating regular motion. The corresponding 3D invariant torus is shown in Figure 11b. Fig. 11b depicts a well formed double-ring shaped 3D invariant torus. Figure 11c depicts the $S(w)$ spectrum for the 1:2 resonant 3D orbit with initial conditions $\left(x_0, p_{x0}\right)$ as in Fig. 3a, while in Figure 11d we see the two 3D islands of invariant tori in the $\left(x, p_x, z\right)$ phase subspace. In Figure 11e we present the $S(w)$ spectrum for the 2:3 resonant 3D orbit with initial conditions $\left(x_0, p_{x0}\right)$ as in Fig. 3a. In this case, we observe three well defined $U$ type structures, as much is the number of 3D islands of invariant tori shown in Figure 11f. Figure 11g and 11h are similar to Fig. 11e and 11f but for the 3:4 resonant 3D orbit with initial conditions $\left(x_0, p_{x0}\right)$ as in Fig. 3a producing four 3D islands of invariant tori. Similarly, Figure 12a illustrates the $S(w)$ spectrum for the 3:5 resonant 3D orbit with initial conditions $\left(x_0, p_{x0}\right)$ as in Fig. 3a, while in Figure 12b the corresponding five 3D islands of invariant tori are presented. In Figure 12c we see the $S(w)$ spectrum for the 4:5 resonant 3D orbit with initial conditions $\left(x_0, p_{x0}\right)$ as in Fig. 3a and in Figure 12d the five 3D islands of invariant tori in the $\left(x, p_x, z\right)$ phase subspace. Furthermore, Figure 12e depicts the $S(w)$ dynamical spectrum for the 5:7 resonant 3D orbit with initial conditions $\left(x_0, p_{x0}\right)$ as in Fig. 3a, while in Figure 12f we see the set of seven small 3D islands of invariant tori. Finally, in Figure 12g one may observe the structure of the $S(w)$ spectrum for the 5:8 resonant 3D orbit with initial conditions $\left(x_0, p_{x0}\right)$ as in Fig. 3a. The corresponding set of eight 3D invariant tori is given in Figure 12h.

For all the $S(w)$ spectrums presented in Figs. 11 an 12 the initial conditions $\left(x_0, p_{x0}\right)$ are as in Fig. 3a, $y_0=p_{z0}=0$, the value of $p_{y0}$ was found always from the energy integral (6), while the initial value of $z_0$ is 0.005. The integration time for the $S(w)$ spectra shown in Figs. 11 and 12 is 5000 time units, while for the 3D invariant tori is equal to $2 \times 10^4$ time units. From the above results, we may conclude that the $S(w)$ spectrum has the ability to identify every type of resonant 3D orbit, since it produces always as many $U$ type structures as much as the total number of 3D islands of invariant tori in the $\left(x, p_x, z\right)$ phase subspace. Moreover, it provides fast and very reliable results not only for simple resonant 3D orbits, but also for complicated resonant orbits of higher multiplicity which correspond to sets of multiple 3D islands of invariant tori.
\begin{figure*}
\centering
\resizebox{0.75\hsize}{!}{\rotatebox{0}{\includegraphics*{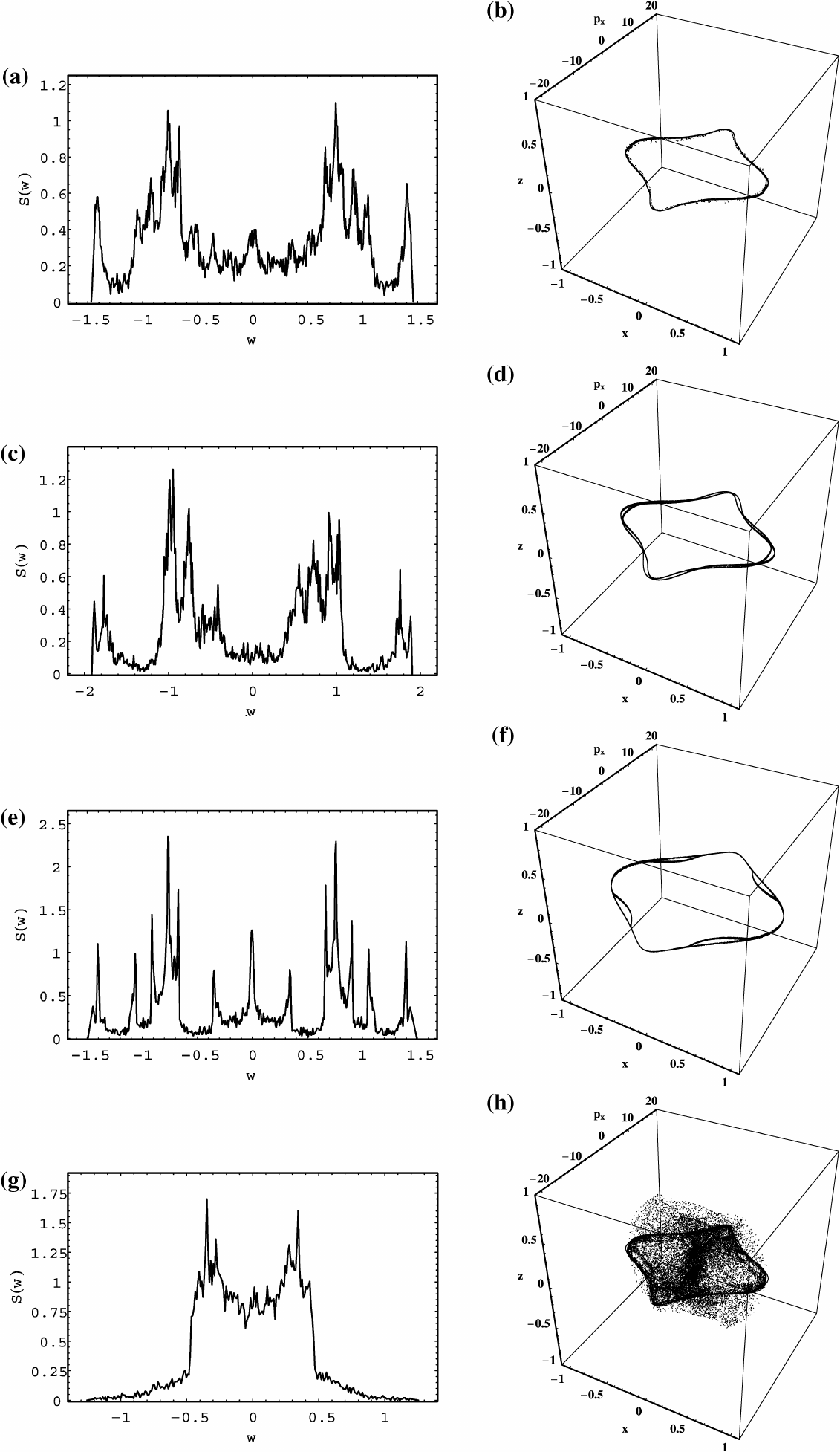}}}
\vskip 0.01cm
\caption{(a-h): (\textit{left pattern}): The $S(w)$ dynamical spectrum for the different chaotic components and the chaotic sea and (\textit{right pattern}) the corresponding $\left(x, p_x, z\right)$ projections on the phase subspace.}
\end{figure*}

In Paper I, we have shown that the different chaotic components in the 3D dynamical system do not merge in order to form a unified chaotic component but instead they continue to exist separately for time intervals much larger than the age of the Universe. Therefore, our next step, would be to deploy the $S(w)$ spectrum in order to identify the several chaotic components of the 3D dynamical system. Figure 13a shows the $S(w)$ spectrum for the first chaotic component (CC-1). We observe a complicated, highly asymmetric structure, with a lot of large, small and abrupt peaks, indicating chaotic motion. The 3D toroidal structure of the first chaotic component (CC-1) is shown in Figure 13b. Similarly, Figure 13c depicts the $S(w)$ spectrum for the second chaotic component (CC-2), while the 3D toroidal structure of CC-2 is presented in Figure 13d. Moreover, Figure 13e depicts the $S(w)$ dynamical spectrum for the third chaotic component (CC-3), while the 3D toroidal structure of CC-3 is presented in Figure 13f. In Figure 13g one may observe the $S(w)$ spectrum for the unified chaotic sea. Things are quite different in the $\left(x, p_x, z\right)$ projection shown in Figure 13h. Here, instead of a well defined toroidal structure, we observe a vague cloudy structure of random 3D points. This cloudy structure indicates that the corresponding 3D orbit is highly chaotic. Here, we must point out that Figs. 13a, 13c, 13e and 13g reveal the sensitivity of the $S(w)$ spectrum. It is obvious that the structure of the $S(w)$ spectrum shown in Fig. 13g is quite different from those shown in Figs. 13a, 13c and 13e. In other words, the $S(w)$ spectrum is sensitive enough allowing us to know whether an 3D orbit corresponds to a chaotic sea or a chaotic component. The initial conditions $\left(x_0, p_{x0}\right)$ for the $S(w)$ spectra corresponding to the chaotic components and the chaotic sea, shown in Fig. 13 are as in Fig. 3b, while the initial value of $z_0$ is 0.01. The integration time for the $S(w)$ spectra shown in Fig. 13 is 5000 time units, while for the 3D  $\left(x, p_x, z\right)$ projections is equal to $5 \times 10^4$ time units.

Our final step, is to verify if the $S(w)$ spectrum can provide reliable results regarding the evolution of 3D sticky orbits. Extensive numerical calculations indicate that unfortunately, the $S(w)$ dynamical spectrum is not sensitive enough in order to provide reliable results about the evolution of 3D sticky orbits. This fact led us in Paper I, to introduce the $S(k)$ spectrum for the study of the 3D sticky motion. The $S(k)$ dynamical spectrum is a modified version of the $S(w)$ spectrum and it was designed especially for the exploration of the 3D sticky motion. Chaotic orbits can be very important in the self-consistency of a dynamical model, especially those being sticky to stable resonant structures for long time intervals (i.e $t > 100$ time units). These sticky orbits appear at the boundaries of transition from an island of stability to the surrounding chaotic zone.
\begin{figure}[!tH]
\centering
\resizebox{\hsize}{!}{\rotatebox{0}{\includegraphics*{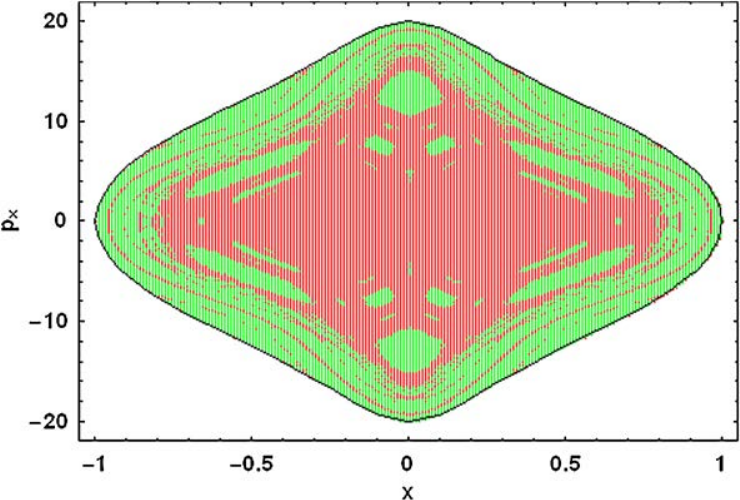}}}
\caption{A grid of initial conditions $(x_0, p_{x0})$ when $z_0=0.007$ constructed using the $S(w)$ spectrum. Inside the \textit{green regions} the 3D motion is ordered, while inside the \textit{red regions} the 3D motion is chaotic. The outermost \textit{black solid line} defines the equation $f\left(x,p_x;z_0\right) = h$. (Color figure online)}
\end{figure}

So far, we have seen that the $S(w)$ dynamical spectrum can provide very reliable results when we study the character of the orbits in a 3D Hamiltonian system. It has the ability to identify complicated resonant 3D orbits and different chaotic components but it can not be applied in order to study the evolution of 3D sticky orbits. However, all the above results obtained from the $S(w)$ are qualitative. Therefore, we must check each time the shape of the spectrum's structure by eye, in order to characterize a 3D orbit. Thus, as we did in the previous section for the $S(g)$ spectrum, we need to establish a numerical criterion, in order to quantify the results provided by the $S(w)$ spectrum. Once more, this criterion can be derived by looking the structure of the spectra shown in Figs. 11, 12 and 13. One may observe, that when the orbit is regular the maximum value of $S(w)$, that is $S(w)_{max}$, is high $\left(S(w)_{max} \geq 7 \right)$. On the contrary, when the orbit is chaotic $S(w)_{max}$ is significantly smaller $\left(S(w)_{max} \leq 3 \right)$. By testing a large number of 3D orbits (approximately 1000), we conclude that when $S(w)_{max} \geq 3.5$ the corresponding 3D orbit is regular, while when $S(w)_{max} < 3.5$ the 3D orbit is chaotic. By using the above criterion for 3D orbits with initial conditions $(x_0, p_{x0})$ on the entire phase plane of Fig. 1 with an initial value $z_0=0.007$ and giving to each initial condition a color according to the value of the $S(w)_{max}$, we can have a clear picture of the regions where chaotic or ordered 3D motion occurs. The result of this procedure for the 3D dynamical system (6), using a dense grid of initial conditions $(x_0, p_{x0}, z_0=0.007)$, is presented in Figure 14. Thus, in Fig. 14 we clearly distinguish between green regions, where the motion is ordered and red regions, where it is chaotic. The outermost black solid line defines the equation
\begin{equation}
f\left(x,p_x;z_0\right) = \frac{1}{2} p_x^2 + V_{tot}(x;z_0) = h.
\end{equation}
Although each 3D orbit in Fig. 14 was computed for only 1000 time units, this time interval was sufficient enough for the clear revelation of the dynamical structure of the 3D phase subspace, when $z=0.007$. The structure of the grid shown in Fig. 14 has a major difference from the grid shown in Fig. 9 corresponding to the 2D model. We observe in Fig. 14 that the two large central regions of stability shown in Fig. 9, which correspond to the 1:1 resonant orbits now have disappeared in the 3D system. This was expected because as we have shown in Paper I, 3D orbits with initial conditions $(x_0, p_{x0})$ inside these islands remain regular only when $z_0 \leq 0.0065$. For all permissible larger initial values of $z_0$ the 3D orbits become chaotic. In Fig. 14 the initial value of $z_0$ is 0.007. Thus, we verified not only the effectiveness but also the reliability of the $S(w)$ spectrum. For a grid of about 6 $\times$ $10^4$ equally spaced initial conditions $(x_0, p_{x0}, z_0=0.007)$, we need about 7.5 hours of CPU time on a Pentium Dual-Core 2.2GHz PC, in order to construct the grid presented in Fig. 14.

\section{Discussion and conclusions}

In the present paper, we applied two different types of dynamical spectra in order to study the properties of motion in Hamiltonian systems of two (2D) and three (3D) degrees of freedom. Our main objective, was to test, verify and prove the effectiveness and also the reliability of these new indicators. For our tests, we chose the dynamical system described by the Hamiltonian (6). Our choice can be justified due to the fact that this potential displays a large variety of resonant orbits, different chaotic components and also several sticky regions in both 2D and 3D systems. Therefore, it provide us an excellent opportunity to test and also reveal the great efficiency of our new dynamical parameters.

For the Hamiltonian system of two degrees of freedom (2D), we used the $S(g)$ dynamical spectrum. The main conclusions of our numerical calculations in the 2D system can be summarized as following:

\textbf{1}. The $S(g)$ dynamical spectrum has the ability to identify any type of complicated resonant 2D orbits, as it produces as many $U$-shaped structures as the total number of the islands of invariant curves in the $\left(x, p_x\right)$ phase plane. Moreover, the numerical outcomes indicate that the $S(g)$ spectrum can identify islandic motion corresponding to higher multiplicity orbits which produce sets of several islands of invariant curves in the phase plane, no matter how tiny these islands are.

\textbf{2}. In a 2D dynamical system where different chaotic components do exist, the $S(g)$ dynamical spectrum is an excellent indicator in order to distinguish between chaotic components and unified chaotic domains. In particular, the structure of the $S(g)$ spectrum corresponding to a chaotic component has many differences from the structure of the $S(g)$ spectrum of a highly chaotic orbit. Therefore, the $S(g)$ spectrum is sensitive enough regarding the chaoticity of a 2D orbit.

\textbf{3}. One more advantage of the $S(g)$ spectrum, is that it can be deployed in order to calculate the sticky period of a 2D sticky orbit and also to follow its time evolution toward the chaotic sea. In the transition regions from ordered to chaotic motion, the pattern of the $S(g)$ spectrum has a small number of abrupt peaks near stable periodic orbits. However, it develops a large number of peaks near the KAM boundaries separating them from the surrounding unified chaotic domain. The $S(g)$ spectra of chaotic orbits near, but outside the KAM boundaries have also a large number of peaks similar to the spectra of orbits near, but inside the same KAM boundaries (see Figs. 7 and 8).

In the case of the Hamiltonian system of three degrees of freedom (3D), we applied the $S(w)$ dynamical spectrum. Our main conclusions of the numerical experiments in the 3D system can be summarized as following:

\textbf{1}. The $S(w)$ dynamical spectrum has the ability to identify any type of complicated resonant 3D orbits, as it produces as many $U$-shaped structures as the total number of the islands of 3D invariant tori in the $\left(x, p_x, z\right)$ projection of the phase subspace. Furthermore, our numerical outcomes indicate that the $S(w)$ spectrum can identify 3D islandic motion corresponding to higher multiplicity orbits which produce sets of islands of 3D invariant tori in the phase subspace, no matter how tiny these 3D islands are.

\textbf{2}. In a 3D dynamical system where different chaotic components still do exist, the $S(w)$ dynamical spectrum is an appropriate indicator in order to distinguish between chaotic components and unified chaotic domains. In particular, the structure of the $S(w)$ spectrum corresponding to a chaotic component has many differences from the structure of the $S(w)$ spectrum of a highly chaotic 3D orbit. Therefore, the $S(w)$ spectrum is sensitive enough regarding the chaoticity of a 3D orbit.

\textbf{3}. Our numerical calculations indicate that unfortunately, the $S(w)$ dynamical spectrum is not sensitive enough in order to provide reliable results about the evolution of 3D sticky orbits. This fact led us in Paper I, to introduce the $S(k)$ spectrum for the study of the 3D sticky motion. Here we must point out, that the $S(k)$ spectrum is a modified version of the $S(w)$ spectrum and it was constructed especially for the study of 3D sticky motion.
\begin{figure*}
\centering
\resizebox{\hsize}{!}{\rotatebox{0}{\includegraphics*{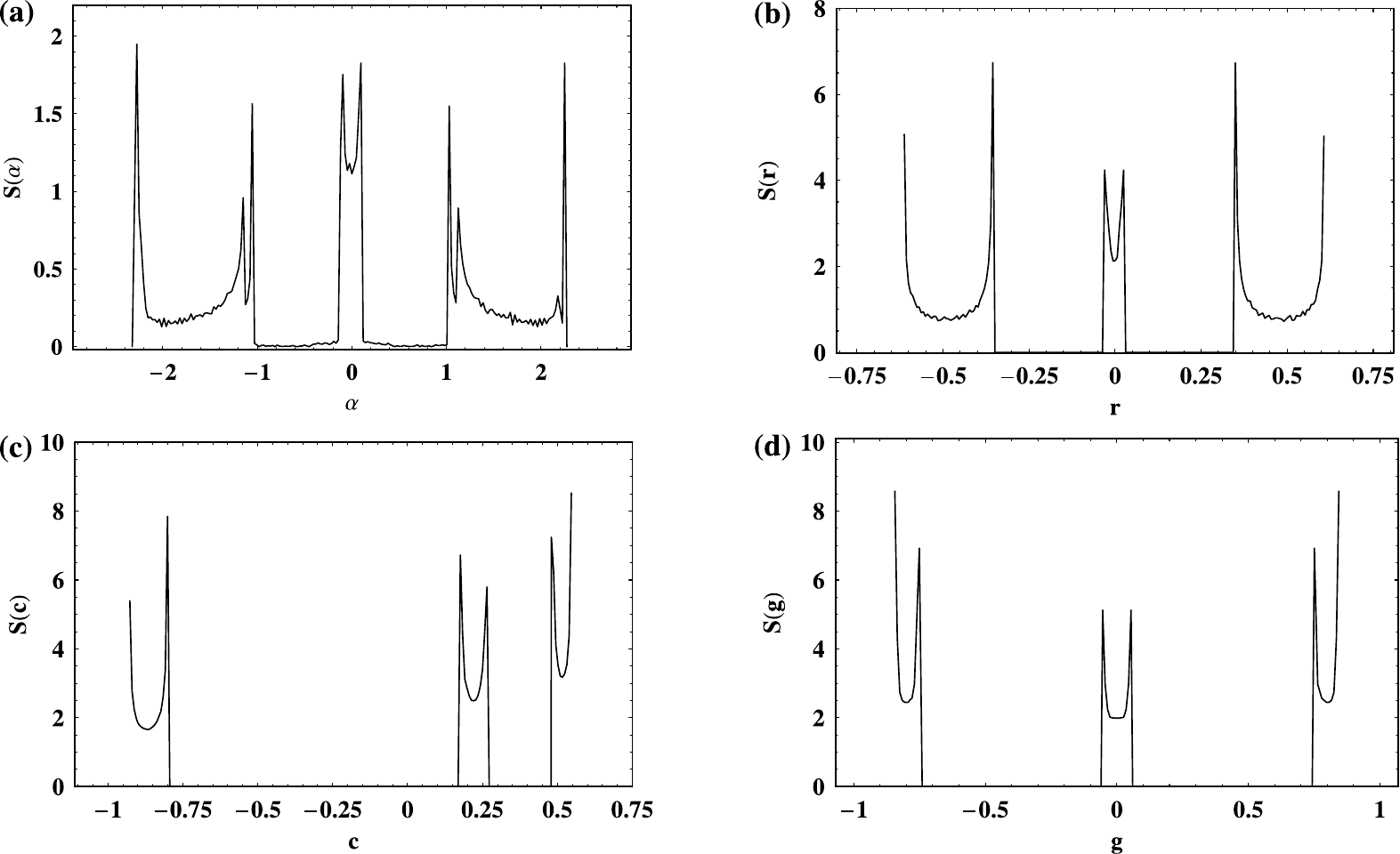}}}
\vskip 0.01cm
\caption{(a-d): (a-\textit{upper left}): The $S(\alpha)$ dynamical spectrum, (b-\textit{upper right}): the $S(r)$ dynamical spectrum, (c-\textit{lower left}): The $S(c)$ dynamical spectrum and (d-\textit{lower right}): The $S(g)$ dynamical spectrum for a 2D orbit. Details are provided in the text.}
\end{figure*}

Thus, taking into account all the above findings, we conclude that the $S(g)$ and $S(w)$ spectrums proved to be very reliable and fast indicators which can help us to identify the character of orbits in Hamiltonian systems. Both spectra need about 5000 time units of integration time, in order to reveal the true nature of an orbit. Moreover, the definitions of these spectra are based on a complicated combination of the coordinates and the momenta (orbital elements), yielding more information about the structure of the orbit. Both spectrum definitions have the value of $p_y$ in their denominators. As the value of $p_y$ is always positive $(p_y > 0)$, while $x$, $z$, $p_x$ and $p_z$ can obtain positive and negative values, we can check the symmetry and locate the position of the corresponding 2D islands of invariant curves in the $\left(x, p_x\right)$ phase plane or the 3D invariant tori in the $\left(x, p_x, z\right)$ phase subspace. In addition, the fact that $p_y$ is always positive is a great advantage because we avoid cases such as $p_y=0$ which would cause problems in the definitions of the $g$ and $w$ parameters.

At this point, it would be of particular interest to compare our new spectrums with previously used spectra. The $S(\alpha)$ spectrum [37] has very limited ability to monitor the evolution of sticky motion. Furthermore, the $S(\alpha)$ spectrum needs the simultaneously computation of two neighboring orbits, while our new spectra do not have this requirement. Another weakness of the $S(\alpha)$ spectrum is that it cannot distinguish complicated resonant orbits producing multiple sets of islands of invariant curves. This is very important because (i) the detection of sets of small islands of invariant curves is significant in order to search for higher order resonances and (ii) there are cases where we have sets of small islands of invariant curves embedded in a large and unified chaotic area. In these cases we must be able to detect the small islands of stability which exist as landmarks of ordered motion. Moreover, the $S(c)$ spectrum [3] in several cases produces less $U$ shaped structures than the total number of the islands of invariant curves or the invariant tori. Note, the coupling of the third component $z$, carrying all the information about the 3D motion, is hidden in the definition of the $S(c)$ spectrum. Thus, the $S(c)$ spectrum provides implicit results regarding the 3D orbits. On the other hand, using the new $S(w)$ spectrum the outcomes are explicit, since the coupling of the third component $z$ is located directly in its definition. Therefore, by introducing the new $S(g)$ and $S(w)$ dynamical spectra, we have managed to overcome all the above drawbacks of the previously used spectra.

It would be very illuminating for the reader, if we display an example comparing the structure of different kinds of dynamical spectra. We will present the dynamical spectra using chronological order. In Figure 15 a-d we observe the structure of four different spectra corresponding to a 2D orbit with initial conditions $x_0=0, y_0=0$ and $p_{x0}=12.5$, while the initial value of $p_y$ is found from the energy integral (8). In Figure 15a we see the $S(\alpha)$ spectrum [37]. It is clear that the $S(\alpha)$ spectrum is quite simple and almost symmetrical with respect to the $\alpha=0$ axis. It tends to become $U$-shaped, but with a lot of small peaks. Moreover, since this orbit produces three islands of invariant curves in the $\left(x, p_x\right)$ phase plane one should also expect three separate $U$-shaped structures. On the contrary, in Fig. 15a we observe that the three spectra are joined together. Figure 15b displays the output of the $S(r)$ dynamical spectrum [20]. Here, we see that the spectrum again is almost symmetrical with respect to the $r=0$ axis. However, with a more closer look one may observe that the $U$-shaped structures shown in Fig. 15b are not very smooth. Furthermore, in some cases the total number of $U$-type structures is smaller than the corresponding number of islands of invariant curves, due to the fact that symmetric islands produce identical spectra. In order to overcome this drawback, a new type of dynamical spectrum was introduced in [3]. In Figure 15c we present the structure of the $S(c)$ spectrum. In this case, the $U$-type structures are very smooth but they are completely asymmetric with respect to the $c=0$ axis. Indeed, the main disadvantage of the $S(c)$ spectrum is its inadequateness to reveal the relative positions of the corresponding islands of invariant curves. In addition, it does not provide reliable results when the orbit is complicated and corresponds to a set of several small islands of invariant curves. The structure of the new $S(g)$ spectrum is shown in Figure 15d. It is obvious, that here everything are as they should have been. The $U$-type structures are separate, very smooth and completely symmetrical with respect to the $g=0$ axis. Thus, one may conclude that the $S(g)$ spectrum is a powerful tool with impressive performance, when it is applied to detect resonant orbits of higher multiplicity. The integration type for all the spectra presented in Fig. 15a-d is 5000 time units.

There are also other dynamical indicators, such as the LCE, FLI [17], SALI [35] and GALI [36], which can be used in order to characterize an orbit. The LCE is a well known dynamical indicator, but it requires vast time intervals of numerical integration of order of $10^5$ time units in order to provide reliable and conclusive results about the nature of an orbit. On the contrary, FLI, SALI and GALI need at most 500 to 1000 time units of numerical integration time, in order to reveal the true nature of an orbit in Hamiltonian systems. The above methods can not be used in order to identify complicated resonant orbits of higher multiplicity. Moreover, FLI and GALI are not very sensitive in the case of sticky motion. However the FLI method is more sensitive in the case of thin chaotic layers. Furthermore, the SALI method can provide reliable information about sticky orbits, using shorter integration time than the $S(g)$ or $S(w)$ spectrums. The main disadvantage of FLI, SALI and GALI is that these indicators need the simultaneously integration of several sets of variational equations and deviations vectors, while the $S(g)$ and $S(w)$ dynamical spectra need only the integration of the basic equations of motion.

All the different spectral definitions used in previous studies could obtain only qualitative results regarding the nature of an orbit. Thus, every time we needed to inspect the shape of the spectrum's structure by eye in order to characterize an orbit. In the present research we have established numerical criteria in order to quantify the results obtained from the $S(g)$ and $S(w)$ dynamical spectra. It was shown in Figs. 9 and 14 that these numerical criteria are valid and therefore, using the $S(g)$ and $S(w)$ dynamical spectra we can reveal the dynamical structure of the phase plane through the construction of a dense grid. In addition, the $S(k)$ spectrum is a very effective tool for the study of 3D sticky motion. Thus, it is the author's humble opinion, that by using the triplet of $\left(S(g), S(w), S(k) \right)$ dynamical spectra, one can study and obtain very fast and reliable results, regarding the structure of Hamiltonian systems of two or three degrees of freedom. Since these new dynamical spectra have been proved extremely accurate and reliable, they may not be amenable to further improvements and thus, we may have marked the end of an era for the dynamical spectra of orbits in Hamiltonian systems.

\section*{Acknowledgments}

The author would like to thank the two anonymous referees for the careful reading of the manuscript and also for their very aptly and creative suggestions and comments which allowed us to improve the quality and the clarity of the present article.

\end{document}